\documentclass[twocolumn,twoside,slac_two]{revtex4}
\pdfoutput=1
\usepackage{graphicx}
\usepackage{fancyhdr}
\begin{document}

\title{\bf Analysis of one hadron rich event}


\author{S.L.C.Barroso}
\affiliation{Departamento de Ci\^encias Exatas/UESB, 45083-900 Vit\'{o}ria da Conquista, BA}

\author{A.O.deCarvalho, J.A.Chinellato, A.Mariano, E.J.T.Manganote, E.C.F.P.Vicente, E.H.Shibuya }
\affiliation{Instituto de F\'{\i}sica `Gleb Wataghin'/UNICAMP, 
13083-859 Campinas, SP}

\begin{abstract}
In this report arguments  are presented  to classify this hadron rich event as an interaction event and the consequences of this statement.  For instance the total invariant mass would be estimated as $\approx$ 61 GeV/c$^{2}$ and the pair of hadrons used for height estimation have invariant mass = 2.2 GeV/c$^{2}$. Besides, tables showing the parametric and non-parametric analysis resulting in a criteria table and the resulting tables for the discrimination of $\gamma$ or hadron induced showers were presented at the $16^{th}$ ISVHECRI, held at Batavia, USA. The main point of hadron rich and Centauro events is the identification of the nature of the observed showers. The identification and energy determination of $\gamma$ or hadron induced showers was made using 2 simulations. Complemented with the observation of photosensitive material under microscope it was determined that the event C16S086I037 could be classified as a hadron rich event. We used 10 reasonable scenarios for $\gamma$/hadron discrimination and obtained that the event is composed of 25 $\gamma$'s, 36 hadrons and 1 surviving and leading hadron. All these scenarios were reported at the $14^{th}$ ISVHECRI, held in Weihai, China and resulted in rather constant values of physical quantities, like the mean transverse momentum of hadrons,  $<P_{T_{h}}>$, and the mean inelasticity of $\gamma$-ray, $<k_{\gamma}>$. Assuming that the most energetic shower is the surviving particle of an interaction and the tertiary produced particles are from normal multiple pion production, the characteristics of the interaction are: Energy of primary particle $E_{0} = 1,061$ TeV, Inelasticity of collision K = 0.81, Mean inelasticity of $\gamma$-ray $<k_{\gamma}>  = 0.27$, Hadron induced showers energy/Total energy
$Q'_{h} = 0.90$ $\approx Q_{h} = 0.71$, Rapidity density $N_{h}/\Delta Y = 
 (8.56-9.89)$, Mean energy of secondary hadrons $<E_{h}> = (21.5\pm4)$ TeV,
Mean transverse momentum $<P_{T_{h}}> = (1.2\pm0.2)$ GeV/c, Upper bound of
partial cross section $\sigma \leq (0.32-0.85) mb $ and life time $\tau \leq
10^{-12}$ s.
\end{abstract}

\maketitle

\thispagestyle{fancy}


\section{Introduction}
The Brazil-Japan Collaboration of the Chacaltaya emulsion chamber experiment (B-J Collaboration) exposed 25 cosmic-ray particle detectors at 540 g/cm$^{2}$ level, geomagnetic coordinates $4^{0}50^{'}40^{''}$ South and $0^{0}50^{'}20^{''}$ East. These detectors consist of multi-layered envelopes containing typically 2 or 3 X-ray films and 1 nuclear emulsion plate, inside a barrier bag, all having an area of 40cm x 50cm and thickness of 200 $\mu$m and 1,550 $\mu$m, respectively \cite{luk93}. The envelopes are inserted between lead plates and the last 11 chambers have two-storey structure as a main detector. 

Since the observation of an unusual event of cosmic ray interaction, nicknamed as a Centauro event, attempts to explain it were under way.
The pioneer event showed  different behaviour from usual interaction events, that is, it shows more particles in the lower part of the detector, hence the nickname. Consequently an empirical interpretation was that it produces charged particles, without $\pi^{0}$s, for instance. After the observation of this event, a search for new Centauro candidates was carried out in later experiments. Unfortunately none of these candidates  showed the visual aspect of the pioneer one, that is upper block of photosensitive detectors less fired than the lower block, but one of these candidates, hereafter nicknamed Centauro V, presented a most relevant feature, with a mean transverse momentum of charged hadrons, $<P_{T_{h}}>$ $\approx 1.0$ GeV/c.

The showers of the first event seen only in the lower chamber were classified as hadrons, due to the small probability %(exp[-8]) 
($e^{-8}$) 
to be of electromagnetic origin. The same feature in another event (C22I019) was observed. The notation used here means that the concerned chamber is C22 (exposed during the period April/86-May/88 and I019 identifies the lower chamber block no.019. In the case of this event, the identification of hadrons is much  clearer due to the absence of showers in the upper chamber. Most other hadron rich events do not have this remarkable feature. Therefore a crucial aspect is the identification of hadrons and some other criteria was used, with emphasis in a comparison with simulated showers. The B-J Collaboration classifies events as hadron-rich for those that have more than 50\% of total observed energy in hadronic origin showers. Previously, \cite{bar02}, the B-J Collaboration showed that there are only 2 other hadron-rich events compatible with the pioneer one (C15S055I012, exposed during the period October/69-July/70). The most straighforward way of knowing the interaction height of an atmospheric event is a geometrical way to measure changes in relative distance of cascade showers along the depth, called the triangulation method. This method was succesfully used in one of these events, C16S086I037, because it had 7 hadronic induced showers with coincidence of showers in both parts of chamber, showing unambiguously the continuation of showers. Moreover all showers are consistently produced in one interaction point, as could be verified using the algorithms reported at ICRC2001 \cite{aug01}.

\section{Description of the event C16S086I037}
This event was observed in the $16^{th}$ chamber of the experiment exposed during the period March/71 - April/72 (383 days of exposure time) and  15+7 events were observed with total electromagnetic energies $\Sigma E_{\gamma} \ge$ 23 TeV and $\Sigma E_{\gamma} \ge$ 20 TeV, respectively and the 7 events are observed
only in the part of chamber below the carbon target. This chamber has an  optimized solid angle design and its thickness is $\approx$ 1.9 $\lambda$, therefore a deep starting shower may be detected. At the location of this event were (3+11) Nuclear Emulsion plates, (6+15) RR-type and (12+31) N-type X-ray films, inserted between lead plates (12+20) r.l. The numbers mean the quantity in ($86^{th}$ upper + $37^{th}$ lower) blocks, respectively. In total it presents 65 showers (numbered \#1 to \#66, neglecting \#14 that has a different azimuthal angle), the zenith angle $\theta = 12.6^{0}$ and the azimuth is $\phi =30 ^{0}$ incident from Northeastern to Southwest direction. As some pairs of showers are close enough and sufficiently isolated from other showers, to analyse this event we considered 6 showers amalgamated into 3 clusters (\#4+\#5, \#43+\#44 and \#47+\#55).  Therefore the total of individual showers analyzed is 62, distributed as ((45-2)+(7-1)+13)) in upper, penetrating from upper to lower and observed only in lower parts of the chamber, respectively. The 7 showers (\#8,\#10,\#11,\#20,\#27,\#37 and \#40) traversing from upper to lower detectors and the 13 showers (\#54 to \#66) observed only after traversing all the upper part, that is after more than 12 r.l., were interpreted as hadronic in origin. The pair (\#47+\#55) is observed in (upper+lower) parts of the chamber, and so interpretable as clusters of one secondary particle of the interaction.  Identification of showers using the Nuclear Emulsion plates, both types of X-ray films and best fitting procedures obtained from simulations were reported already in some conferences and symposia, the last one at the $14^{th}$ ISVHECRI, held in Weihai, China \cite{bar08}      

From 2 showers (\#20 and \#27), both showing (5upper+2lower) cores and (5upper+3lower) cores we determined the interaction point as H=($478^{+188}_{-106}$)m at the X-ray RR-type films and H=($500^{+206}_{-113}$) m in nuclear emulsion plates. Considering 4 paired combinations of 4 showers, that is avoiding 2 negative relative distances, we determined the mean height H=($412^{+327}_{-126}$) m in nuclear emulsion plates. Considering all 6 combinations and the absolute values of relative distances, the mean height we obtained is H=($552^{+199}_{-126}$) m, so that the value of the interaction height seems to span the interval [300,750 m].  

\begin{figure}
\includegraphics[width=95mm]{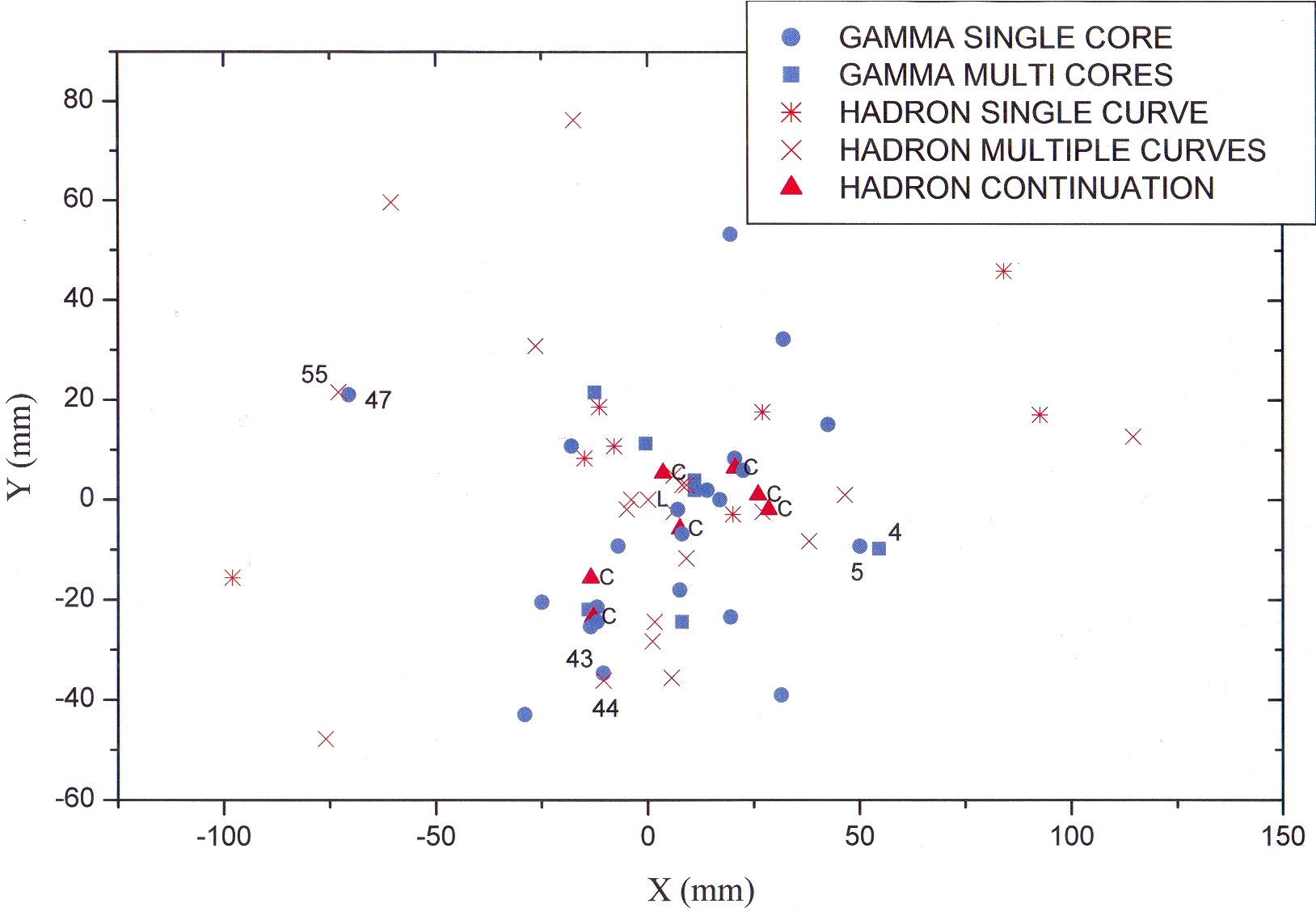}
\caption{Map of event C16S086I037}
\label{Fig.1}
\end{figure}

Figure \ref{Fig.1} shows the map of C16S086I037 projected in a plane perpendicular to their incident direction and the showers amalgamated into 3 clusters are identified.

\section{Analysis results}  
Energy determination was made using a simulation carried out by T.Shibata et al.\cite{shi87,shi89} and M.Tamada\cite{tam00}, this one using the Corsika computer code  with a QGS-jet interaction model applied only for the upper chamber blocks. Another use of these simulations is to identify hadronic showers through the best fitting procedure that provides statistical parameters like $\sigma$ and $\chi^{2}$.
Some cluster showers are better fitted using double transition curves,
therefore these showers may be identified as hadron induced ones. To identify
these showers as hadron induced showers more confidently, another criterion such as the observation of multi-core structure was used.

For single showers the main tool for $\gamma$ or hadron induced shower recognition is the comparison with simulated events, using $\sigma$ and $\chi^{2}$  fitting to the data of darkness of the showers in the X-ray films. Some of the identified hadrons by Shibata's simulation have the aspect of double peaks in the shower development inside the chamber. As even a $\gamma$-shower may have this kind of development, through the Landau-Pomeranchuk-Migdal effect, in spite of $\sigma_{min}$ for these showers, we complemented the before mentioned statistical parameters with other criteria such as observation of multicored structure in Nuclear Emulsion plates and constructed a score table from which we selected the $\gamma$-hadron samples which have at least two of the adopted criteria,  the most important being both statistical parameters.

To classify this event as the product of interaction and not of fragmentation  we used the  before mentioned algorithms \cite{aug01}, obtaining: 

\noindent
i) A correlation   
\begin{equation}
R = \frac{[\Sigma E_{i}][\Sigma E_{i}(\Gamma\theta_{i})^{2}]}{[\frac{4}{\pi}
\Sigma E_{i}(\Gamma\theta_{i})]^2}
\end{equation}

\noindent
ii) From the correlation 
\begin{eqnarray}
mDW = \frac {1}{4M\Gamma}[\Sigma E_{i} + \frac{4}{\pi}\Sigma E_{i}
(\Gamma\theta_{i}) + \Sigma E_{i}(\Gamma\theta_{i})^{2} + \nonumber \\ 
\frac{4}{3\pi} \Sigma E_{i}(\Gamma\theta_{i})^{3}]
\end{eqnarray}
we obtained the angular coeficient s=1.66, rather different from 2.0 that is for isotropic decay of the secondaries of the interaction.

\noindent iii) A correlation, known as the Peyrou plot \cite{pey64} shows a big dispersion in the longitudinal, but not in the transverse momentum, even calculated from Center of Mass of only hadrons or Center of Mass of all showers.

The shower (\#62) was not observed in the upper block and it interacts twice
in the lower chamber. It has 16\% of the total energy much more
than the second highest energy shower (\#27) that has 8\%, and so we can legitimately identify it as the surviving hadron.  
This identification
permits to calculate K, $<k_{\gamma}>$, $E_{0}$ and $Q'_{h}$, that is, the
inelasticity of the nuclear collision, the $\gamma$-ray inelasticity, the energy
of the primary particle and the fraction of hadron-induced showers energy to total energy.

\begin{figure}[h!t]
	\centering
	\includegraphics[totalheight=8cm,origin=c]{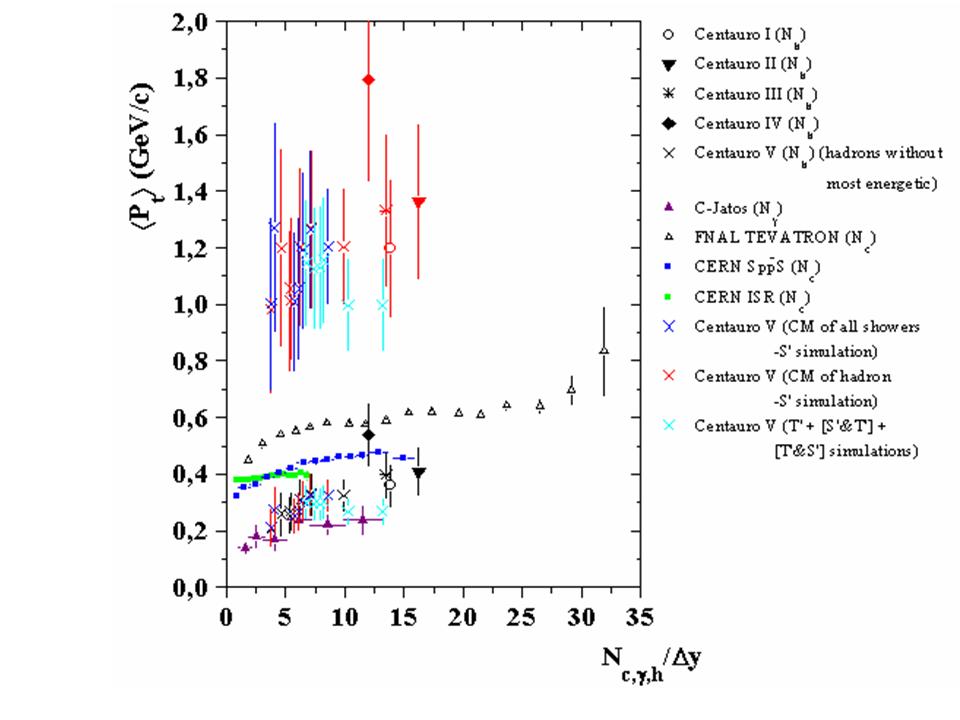}
	\caption{Mean $P_{T}$ - rapidity density correlation. The cross marks are for this event under various scenarios of hadron identification}
	\label{Fig.2}
\end{figure}

The correlation Mean Transverse Momenta - Rapidity density is shown in  figure \ref{Fig.2} where we 
observe that this event is analysed as not due to $\gamma$ induced shower events, but, considered  effectively as a 
hadronic shower,  is in a region above the data of accelerator experiments. 

Concerning the cross section we made an estimate based on an expression used in accelerator 
experiments \cite{gar87}, assuming that the chamber is a detector of cosmic ray accelerated 
particles. The values obtained are $\sigma_{normal}$=[16-43]mb and $\sigma_{hadron-rich}$=[0.32-0.85]mb.
Assuming the Breit-Wigner cross section for the  interaction p+N $ \rightarrow $ c16s086i037 $ \rightarrow hadrons + \gamma$'s we obtained $\Delta T_{C16s086i037} < 10^{-12}$s for the mean life time. 

\section{Conclusions and discussion}
\noindent i) Four figures presented at the symposium suggest that the event is the product of an interaction or of an unusual fragmentation. An additional suggestion for this interpretation comes from the fact that the difference between the center of momenta of all showers and the center of hadronic showers, neglecting shower \#62, is small ($\Delta r_{max} = 4 mm$ implying in $\Delta \theta_{max} = 8 \times 10^{-6}$ rad), but not zero.

\noindent ii) Two other figures indicate that  shower \# 62 is a leading energy and surviving hadronic induced shower.

\noindent iii) The ratio between the maximum darkness of the first peak (d$_{1}$) to the sum (d$_{1}$ + d$_{2}$) is 0.37 and the corresponding energy ratio is 0.31. Other similar estimations are the energy ratio between surviving hadron to all hadrons = 0.26 and the ratio between surviving hadron energy to total observed shower energy = 0.24. Then we used $<k_{\gamma}> = 1/3$ in the conservation laws of energy and linear momentum and got the values listed in a table presented at the symposium.

\noindent iv) The analysis shows that this event presents high energy content in hadronic showers (Q=0.71)and (Q$^{'}=0.90$), characterizing a hadron-rich event and that it is above the region of accelerator data, as shown in figure \ref{Fig.2}.

\noindent v) Using the interaction height (vertex), we obtained for the invariant mass of the pair of hadrons (\#20 and \#27) the value $2.2 GeV/c^{2}$.  We obtained almost the same value ($2.9 GeV/c^{2}$) assuming that the transverse momentum of the primary cosmic ray particle expresses the invariant mass, an assumption that is correct in a thermodynamical model for interaction.

\noindent vi) The mean height obtained from a kinematical coupling of 2 $\gamma^{'}s$ induced showers to $\pi^{0}$ has the value around 780 m. This height is compatible with the mean height obtained from triangulation of all pairs of hadronic induced showers (764 m). This analysis is preliminary and not yet complete.

\noindent vii) The obtained value for $<k_{\gamma}>= 0.27$ is also consistent with charge independence for Multiple Meson Production, considering the 25$\gamma s$ coupled to 12$\pi^{0}$s and 36$\pi^{\pm}$.

\noindent viii) A comparison of this event with other similar events was made in \cite{bar02}. There it was shown that it is compatible with Centauro I and with Centauro IV. As the particles of Centauro I are spread covering an area of $\approx$1.2 cm of radius ($\approx$ 10\% radius of this event(10.8~cm)), it is plausible that the Centauro I event has its vertex located near to 50 m. above the chamber, as previously shown.

\noindent ix) Centauro IV was analysed by M.Tamada \cite{tam77} showing that the fractional energy spectrum has a slope similar to the corresponding spectrum of Centauro I, so it is similar indirectly with this event and also with the other 4  events containing a halo in the center \cite{slcbar08}.              

Thus, the results of our observations and analysis are compatible with the assumption that this hadron-rich event is the result of a collision of a
hadron with energy $E_{0} \approx 1,000 TeV $ at H=($500^{+206}_{-113}$)
m above the chamber, with mean transverse momenta $<p_{t}> \approx 1 GeV/c$.  That is to say this event is an authentic Centauro event producing $\approx$ 70-90 hadrons at the interaction point.

We dedicate this analysis to pioneers of the Brazil-Japan Collaboration,  Professors Y.Fujimoto, S.Lasegawa and C.M.G.Lattes and homage to our late friend of the Castor experiment, Professor Aris Angelis also an enthusiast of Centauro events.

\section{Acknowledgments}

We have pleasure in acknowledging our indebtedness to the financial
support from CBPF, CNEN, CNPq, FAEPEX-UNICAMP, FAPESP, FINEP in
Brazil, and Aoyama Gakuin University, University of Tokyo, Wasedi
University, Ministry of Education in Japan.  We are also grateful to
IIF-UMSA in Bolivia, host of the experiments, for the help on many
occasions. The authors wish to acknowledge all members of the B-J
Collaboration, for the free use of the data. Two of us (SLCB and EHS)
are keen to express their gratitude to A.Ohsawa, M.Tamada and T.Shibata for the simulated events that we used in this and previous reports. One of us, SLCB,
wish to acknowledge the funding agency FAPESP for support (contract 
no. 01/04150-0).

\end{document}